\pdfoutput=1

\documentclass[sigplan,10pt,screen]{acmart}

\usepackage{xspace}
\usepackage{enumitem}
\usepackage{balance}
\usepackage{xcolor}
\usepackage{booktabs, multirow}
\usepackage{soul}
\usepackage{listings}
\usepackage{subcaption}
\usepackage{wrapfig}
\usepackage[ruled, vlined, linesnumbered, commentsnumbered, longend]{algorithm2e}
\usepackage{tcolorbox}

\begin{document}

\newcommand{\eg}{\emph{e.g.,}\xspace}
\newcommand{\ie}{\emph{i.e.,}\xspace}

\newcommand{\parabf}[1]{\noindent\textbf{#1}}

\newcommand{\CodeIn}[1]{{\small \texttt{#1}}}
\newcommand{\Comment}[1]{}

\newcommand{\lingming}[1]{{\color{red}\bfseries [Lingming: #1]}}
\newcommand{\cyy}[1]{\textcolor{cyan}{[\textit{Chenyuan:} #1]}}
\newcommand{\zijie}[1]{\textcolor{olive}{[\textit{Zijie:} #1]}}
\newcommand{\haoyu}[1]{\textcolor{blue}{[\textit{Haoyu:} #1]}}
\newcommand{\Zichen}[1]{\textcolor{blue}{[\textit{Zichen:} #1]}}

\newcommand{\revision}[1]{{#1}}

\newcommand{\tech}{{KNighter}\xspace}

\newcommand{\crix}{{CRIX}\xspace}
\newcommand{\goshawk}{{Goshawk}\xspace}
\newcommand{\llift}{\textsc{LLift}\xspace}
\newcommand{\vulrag}{\textsc{Vul-RAG}\xspace}
\newcommand{\inferroi}{\textsc{InferROI}\xspace}

\newcommand{\llm}{{LLM}\xspace}
\newcommand{\llvm}{{LLVM}\xspace}

\newcommand{\clang}{{Clang}\xspace}
\newcommand{\csa}{{CSA}\xspace}
\newcommand{\smatch}{{Smatch}\xspace}

\newcommand{\correct}{{plausible}\xspace}

\newcommand{\numValid}{{39}\xspace}

\newcommand{\numEvalBug}{{62}\xspace}
\newcommand{\numNewBug}{{30}\xspace}

\newcommand{\numBug}{{92}\xspace}
\newcommand{\numConfirmed}{{77}\xspace}
\newcommand{\numPending}{{15}\xspace}
\newcommand{\numFixed}{{57}\xspace}
\newcommand{\numCVE}{{30}\xspace}

\newcommand{\avgYear}{{4.3}\xspace}
\newcommand{\overfive}{{26}\xspace}
\newcommand{\overfifteen}{{6}\xspace}

\setlength{\fboxsep}{0.3pt}
\definecolor{evalColor}{HTML}{d7ebff}
\definecolor{npdColor}{HTML}{fdeeff}

\definecolor{codegreen}{rgb}{0,0.6,0}
\definecolor{codegray}{rgb}{0.5,0.5,0.5}
\definecolor{codepurple}{rgb}{0.58,0,0.82}
\definecolor{backcolour}{rgb}{0.97,0.97,0.95}
\definecolor{forestgreen}{rgb}{0.28,0.62,0.37}
\definecolor{codeblue}{rgb}{0,0.5,1}

\definecolor{textbackcolor}{rgb}{0.96,0.96,0.96}
\definecolor{promptbackcolor}{rgb}{0.93,0.93,0.99}

\lstdefinestyle{promptstyle}{
    backgroundcolor=\color{promptbackcolor},   
    numberstyle=\tiny\color{codegray},
    stringstyle=\color{blue},
    basicstyle=\ttfamily\scriptsize,
    breakatwhitespace=false,         
    breaklines=true,                 
    breakindent=0pt,
    captionpos=b,                    
    keepspaces=true,                 
    numbers=none,                    
    numbersep=5pt,                  
    showspaces=false,                
    showstringspaces=false,
    showtabs=false,                  
    tabsize=4,
    keywordstyle={},                 %
}

\lstdefinestyle{textstyle}{
    backgroundcolor=\color{textbackcolor},   
    numberstyle=\tiny\color{codegray},
    stringstyle=\color{blue},
    keywordstyle=none,
    commentstyle=none,
    basicstyle=\ttfamily\scriptsize,
    breakatwhitespace=false,         
    breaklines=true,                 
    breakindent=0pt,
    captionpos=b,                    
    keepspaces=true,                 
    numbers=none,                    
    numbersep=5pt,                  
    showspaces=false,                
    showstringspaces=false,
    showtabs=false,                  
    tabsize=4,
    keywordstyle={},                 %
}

\lstdefinestyle{mystyle}{
    backgroundcolor=\color{backcolour},   
    commentstyle=\color{codepurple},
    keywordstyle=\color{codepurple},
    numberstyle=\tiny\color{codegray},
    stringstyle=\color{blue},
    basicstyle=\ttfamily\scriptsize,
    breakatwhitespace=false,         
    breaklines=true,                 
    captionpos=b,                    
    keepspaces=true,                 
    numbers=left,                    
    numbersep=5pt,                  
    showspaces=false,                
    showstringspaces=false,
    showtabs=false,                  
    tabsize=4,
}

\lstset{style=mystyle}

\lstdefinestyle{ruststyle}{
    morekeywords={
        as, break, const, continue, crate, else, enum, extern, false, fn, for, if, impl, in, let, loop, match, mod, move, mut, pub, ref, return, self, Self, static, struct, super, trait, true, type, unsafe, use, where, while, dyn, abstract, alignof, become, box, do, final, macro, offsetof, override, priv, proc, pure, sizeof, typeof, unsized, virtual, yield, async, await, try
    },
    sensitive=true, %
    morecomment=[l]{//},  %
    morecomment=[s]{/*}{*/},  %
    morestring=[b]",  %
    morestring=[b]{'}, %
    keywordstyle=\color{codepurple},  %
    commentstyle=\color{codegray}\itshape,  %
    stringstyle=\color{blue},  %
    identifierstyle=\color{black},  %
    ndkeywordstyle=\color{purple}\bfseries,  %
    basicstyle=\ttfamily\scriptsize,  %
    showstringspaces=false,  %
    tabsize=4,  %
    breaklines=true,  %
    breakatwhitespace=false,  %
    showtabs=false,  %
    showspaces=false,  %
    showstringspaces=false,  %
    numbers=left,  %
    numberstyle=\tiny\color{gray},  %
}

\lstdefinelanguage{Verus}{
    style=ruststyle,
    morekeywords=[2]{ requires, ensures, invariant, spec, proof, decreases },
    keywordstyle=[2]\color{red}
}

\lstdefinestyle{cppstyle}{
    morekeywords={
        alignas, alignof, and, and_eq, asm, auto, bitand, bitor, bool, break,
        case, catch, char, char8_t, char16_t, char32_t, class, compl, concept,
        const, constexpr, const_cast, continue, co_await, co_return, co_yield,
        decltype, default, delete, do, double, dynamic_cast, else, enum, explicit,
        export, extern, false, float, for, friend, goto, if, inline, int, long,
        mutable, namespace, new, noexcept, not, not_eq, nullptr, operator, or,
        or_eq, private, protected, public, register, reinterpret_cast,
        requires, return, short, signed, sizeof, static, static_assert,
        static_cast, struct, switch, template, this, thread_local, throw,
        true, try, typedef, typeid, typename, union, unsigned, using, virtual,
        void, volatile, wchar_t, while, xor, xor_eq
    },
    sensitive=true, %
    morecomment=[l]{//}, %
    morecomment=[s]{/*}{*/}, %
    morestring=[b]", %
    morestring=[b]', %
    moredirectives={include, define, undef, if, ifdef, ifndef, else, elif,
        endif, line, error, pragma}, %
    keywordstyle=\color{codepurple}, %
    commentstyle=\color{codegray}\itshape, %
    stringstyle=\color{blue}, %
    identifierstyle=\color{black}, %
    ndkeywordstyle=\color{purple}\bfseries, %
    directivestyle=\color{darkgreen}, %
    basicstyle=\ttfamily\scriptsize, %
    showstringspaces=false, %
    tabsize=4, %
    breaklines=true, %
    breakatwhitespace=false, %
    showtabs=false, %
    showspaces=false, %
    numbers=left, %
    numberstyle=\tiny\color{gray}, %
}

\lstset{
    language=C++,
    style=cppstyle
}

\definecolor{codepurple}{rgb}{0.5,0,0.5}
\definecolor{codegray}{rgb}{0.5,0.5,0.5}
\definecolor{darkgreen}{rgb}{0,0.6,0}
\definecolor{bgLightGreen}{rgb}{0.95,1,0.95}
\definecolor{bgLightRed}{rgb}{1,0.95,0.95}

\lstdefinelanguage{diff}{
  morecomment=[f][\color{blue}]{\@\@},
  morecomment=[f][\color{darkgreen}]{+++},
  morecomment=[f][\color{red}]{---},
  morecomment=[l]{\#},
  commentstyle=\color{codegray}\itshape,
  keywordstyle=[1]{\color{red}},
  keywordstyle=[2]{\color{darkgreen}}
}

\lstdefinestyle{diffstyle}{
    language=diff,
    basicstyle=\ttfamily\scriptsize,
    showstringspaces=false,
    tabsize=4,
    breaklines=true,
    breakatwhitespace=false,
    showtabs=false,
    showspaces=false,
    numbers=left,
    numberstyle=\tiny\color{gray},
    columns=flexible,
    keepspaces=true,
    commentstyle=\color{codegray}\itshape,
    keywordstyle=\color{codepurple},
    stringstyle=\color{blue},
    identifierstyle=\color{black}
}

\newcommand{\codedelete}{red!60}

\newcommand{\CodeDelete}{\makebox[0pt][l]{\color{red!20}\rule[-0.45em]{\linewidth}{1.3em}}\textbf{\color{red}-}}
\newcommand{\CodeAdd}{\makebox[0pt][l]{\color{green!20}\rule[-0.45em]{\linewidth}{1.3em}}\textbf{\color{darkgreen}+}}
\newcommand{\CodeBG}{\makebox[0pt][l]{\color{Yellow!50}\rule[-0.45em]{\linewidth}{1.3em}}}

\title{\tech: Transforming Static Analysis with LLM-Synthesized Checkers}

\author{Chenyuan Yang}
\orcid{0000-0002-7976-5086}
\affiliation{%
  \institution{University of Illinois Urbana-Champaign}
  \country{USA}
}
\email{cy54@illinois.edu}

\author{Zijie Zhao}
\orcid{0009-0008-0718-3088}
\affiliation{%
  \institution{University of Illinois Urbana-Champaign}
  \country{USA}
}
\email{zijie4@illinois.edu}

\author{Zichen Xie}
\orcid{0009-0008-3905-3456}
\affiliation{%
  \institution{Zhejiang University}
  \country{China}
}
\email{xiezichen@zju.edu.cn}

\author{Haoyu Li}
\orcid{0000-0003-0084-1718}
\affiliation{%
  \institution{Shanghai Jiao Tong University}
  \country{China}
}
\email{learjet@sjtu.edu.cn}

\author{Lingming Zhang}
\orcid{0000-0001-5175-2702}
\affiliation{%
  \institution{University of Illinois Urbana-Champaign}
  \country{USA}
}
\email{lingming@illinois.edu}

\copyrightyear{2025}
\acmYear{2025}
\setcopyright{cc}
\setcctype{by}
\acmConference[SOSP '25]{ACM SIGOPS 31st Symposium on Operating Systems Principles}{October 13--16, 2025}{Seoul, Republic of Korea}
\acmBooktitle{ACM SIGOPS 31st Symposium on Operating Systems Principles (SOSP '25), October 13--16, 2025, Seoul, Republic of Korea}\acmDOI{10.1145/3731569.3764827}
\acmISBN{979-8-4007-1870-0/2025/10}



\def\thefootnote{\arabic{footnote}}

\begin{abstract}

Static analysis is a powerful technique for bug detection in critical systems like operating system kernels. However, designing and implementing static analyzers is challenging, time-consuming, and typically limited to predefined bug patterns. While large language models (LLMs) have shown promise for static analysis, directly applying them to scan large systems remains impractical due to computational constraints and contextual limitations.

We present \tech, the first approach that unlocks scalable LLM-based static analysis by automatically synthesizing static analyzers from historical bug patterns. Rather than using LLMs to directly analyze massive systems, our key insight is leveraging LLMs to generate specialized static analyzers guided by historical patch knowledge. \tech implements this vision through a multi-stage synthesis pipeline that validates checker correctness against original patches and employs an automated refinement process to iteratively reduce false positives.
Our evaluation on the Linux kernel demonstrates that \tech generates high-precision checkers capable of detecting diverse bug patterns overlooked by existing human-written analyzers.
To date, \tech-synthesized checkers have discovered \numBug new, critical, long-latent bugs (average \avgYear years) in the Linux kernel; \numConfirmed are confirmed, \numFixed fixed, and \numCVE have been assigned CVE numbers.
This work establishes an entirely new paradigm for scalable, reliable, and traceable LLM-based static analysis for real-world systems via checker synthesis.

\end{abstract}
\settopmatter{printfolios=true}
\maketitle

\section{Introduction}
\begin{figure}[]
    \centering
    \includegraphics[width=0.99\linewidth]{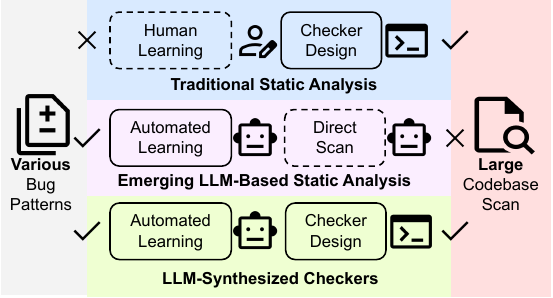}
    \caption{\textbf{Motivation.} Static analysis should scale to address a diverse range of bug patterns and handle massive codebases. \emph{Traditional static analysis} struggles with covering wide-ranging bugs, whereas \emph{LLM-based methods} face hurdles in scaling to large codebases.}
    \label{fig:motivation}
\end{figure}

The reliability of fundamental software systems---particularly operating system (OS) kernels---hinges on robust defect detection methodologies~\cite{kernel-fuzzing,kernelgpt,kernel-static-refcounting,kernel-static-cai2023place,kernel-static-lyu2022goshawk,suzuki2024fitx,yang2024whitefox,gong2025snowplow,gong2023snowcat}. 
Among various techniques, \textit{static analysis}~\cite{ayewah2008using} stands out for its ability to examine source code without execution, making it indispensable for scenarios involving hardware-dependent drivers, complex or rarely exercised paths, and configurations difficult to reproduce in real environments~\cite{kernel-static-lu2019detecting,kernel-static-lyu2022goshawk,kernel-static-refcounting,suzuki2024fitx,gong2025snowplow,gong2023snowcat}. 
Compared to dynamic approaches such as fuzzing---which requires an execution environment and thus only tests actual runtime paths~\cite{syzkaller,kernelgpt, deng2023large,yang2023fuzzing}---static analysis can (in principle) cover all potential execution paths, including corner cases that are seldom triggered in practice. 
While formal verification techniques~\cite{yang2024autoverus,lattuada2023verus,sun2024clover} offer stronger correctness guarantees, their high manual overhead renders them impractical for large-scale systems like OS kernels, making static analysis a more scalable and feasible solution.

\parabf{The static analysis problem.}  
Large-scale systems present a dual challenge for static analysis: addressing diverse bug patterns and managing enormous codebases, as illustrated in Figure~\ref{fig:motivation}. An ideal static analyzer should \emph{(i)} detect a wide range of defects---including those related to nuanced, system-specific semantics---and \emph{(ii)} efficiently process millions of lines of code.
However, existing techniques typically compromise on one of these critical objectives.

\emph{Traditional static analysis.}
Traditional static analyzers are effective in identifying certain bug types, yet they fundamentally rely on pre-defined, rule-based, or formally modeled checks.
This reliance necessitates extensive domain expertise and substantial engineering effort for their development and maintenance~\cite{kernel-static-refcounting,kernel-static-cai2023place,smatch}. Consequently, these tools are often fine-tuned to a narrow subset of bug patterns, which not only limits their ability to detect unforeseen defects but also hampers their scalability in automatically addressing a broader spectrum of issues.

\emph{Emerging LLM-based static analysis.}
On the other hand, Large Language Models (LLMs) are compelling tools for discovering bug patterns in part because they can learn directly from historical patch commits---a treasure trove of real fixes and associated bug contexts~\cite{kernel-static-refcounting,seal,aphp}. Their ability to parse both textual and code content~\cite{achiam2023gpt,team2023gemini,grattafiori2024llama} suggests that LLMs can adapt to new bug types without explicit rule-crafting. However, directly deploying LLMs on large-scale systems (e.g., the Linux kernel at over 30 million lines of code) confronts severe limitations. Their bounded context windows make it impossible to upload all relevant source code at once, and doing so repeatedly would also incur prohibitive computational costs (potentially hundreds of dollars per thorough scan). In addition, LLMs can hallucinate~\cite{huang2025survey,du2024vul,li2024llm}, producing plausible but incorrect outputs, especially when faced with the intricacies of large-scale systems~\cite{mathai2024kgym}.

\parabf{Insight.}  
Can we scale and automate static analysis to handle both diverse bug patterns and enormous codebases? We answer this question by harnessing the strengths of traditional static analysis alongside emerging \llm{s}.
More specifically, we propose \emph{synthesizing static checkers} using \llm{s} rather than applying LLM-based analysis directly to the entire codebase.
In this paradigm, \llm{s} learn bug patterns from historical patches, and these insights are encoded into dedicated static analysis checkers.
This method circumvents the prohibitive costs and context-length limitations of scanning vast codebases while maintaining the flexibility needed to address a wide spectrum of bugs.
Moreover, by validating each synthesized checker against the original patches, we mitigate hallucinations and produce transparent, human-readable logic that developers can trust and maintain.

\parabf{Technical challenges and our solutions.}
Although automated checker synthesis holds significant promise, generating complete static analysis logic remains a formidable challenge---even experts struggle with it. To address this, we introduce a \emph{multi-stage synthesis pipeline} (\S~\ref{subsec:checker-syn}) that decomposes checker generation into manageable subtasks. Furthermore, to enhance the quality of the synthesized checkers by reducing false positives, we develop a \emph{fully automated refinement pipeline} (\S~\ref{subsec:checker-evo}) that leverages bug report triage agents. Together, these pipelines yield checkers that are robust and practical for deployment in real-world scenarios.

We implement our approach in a tool, \tech, the first fully automated pipeline for synthesizing static analyzers, built upon the open-source Clang Static Analyzer (\csa)~\cite{clang-static-analysis}.
While the methodology generalizes to different systems, we target the Linux kernel, one of the most fundamental software systems.
In the evaluation of 61 diverse bug-fix patches, \tech synthesized the high-quality checkers for 61\% of them, achieving a false positive rate of about 35\% aided by the triage agent.
Demonstrating practical impact, \tech has uncovered \numBug new, long-latent vulnerabilities (average \avgYear years) in the Linux kernel, resulting in \numConfirmed developer confirmations, \numFixed fixes, and \numCVE CVEs.
Furthermore, the vulnerabilities detected are orthogonal to those found by existing expert-written analyzers~\cite{smatch}.
These findings validate our approach's efficacy and its contribution to system reliability.

Our main contributions are summarized as follows:

\begin{itemize}[topsep=1pt]
    \item \textbf{Novelty.} We introduce a pioneering approach for synthesizing static analyzers from patch commits. To our knowledge, \tech is the first fully automated static analyzer generation system, establishing a new paradigm for \llm{}-based static analysis.

    \item \textbf{Approach.} We implement \tech with multi-stage synthesis and automated refinement pipelines for the Linux kernel. This design enables detection of diverse bug classes in large-scale systems.

    \item \textbf{Evaluation.} We demonstrate that \tech successfully synthesizes effective checkers from the Linux kernel bug-fix patches across various bug categories, achieving practical false positive rates.

    \item \textbf{Real-world impact.} \tech-generated checkers have discovered \numBug new, long-latent (average \avgYear years) bugs in the Linux kernel, with \numConfirmed confirmed, \numFixed fixed, and \numCVE assigned CVE numbers---demonstrating its practical impact on system reliability and security.
\end{itemize}

\begin{center}
\includegraphics[scale=0.0185]{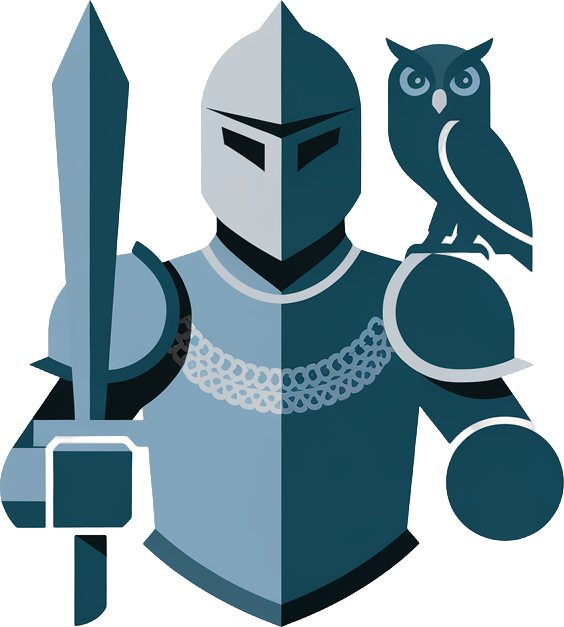} {\tech} is available at \href{https://github.com/ise-uiuc/KNighter}{ise-uiuc/KNighter}.
\end{center}

\section{Background and Motivation}
\begin{figure}[t!]
    \centering
\begin{subfigure}[t]{0.95\linewidth}
\includegraphics[width=\linewidth]{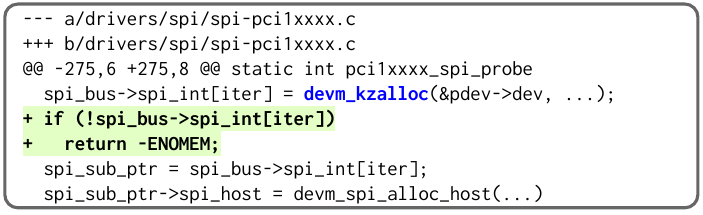}
\caption{Patch for a Null-Pointer-Dereference bug. The pointer returned by \CodeIn{devm\_kzalloc} should be checked.}
\label{fig:kzalloc-patch}
\end{subfigure}

\begin{subfigure}[t]{0.95\linewidth}
\includegraphics[width=\linewidth]{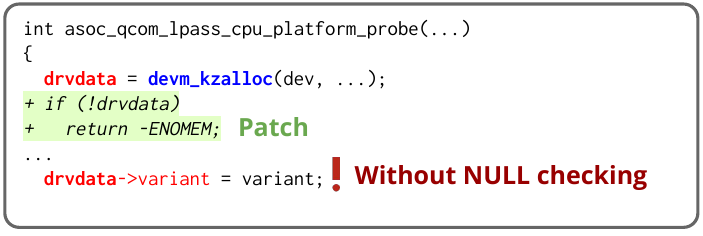}
\caption{A new bug detected by \tech with \textbf{CVE-2024-50103}.}
\label{fig:kzalloc-bug}
\end{subfigure}

\begin{subfigure}[t]{0.95\linewidth}
\includegraphics[width=\linewidth]{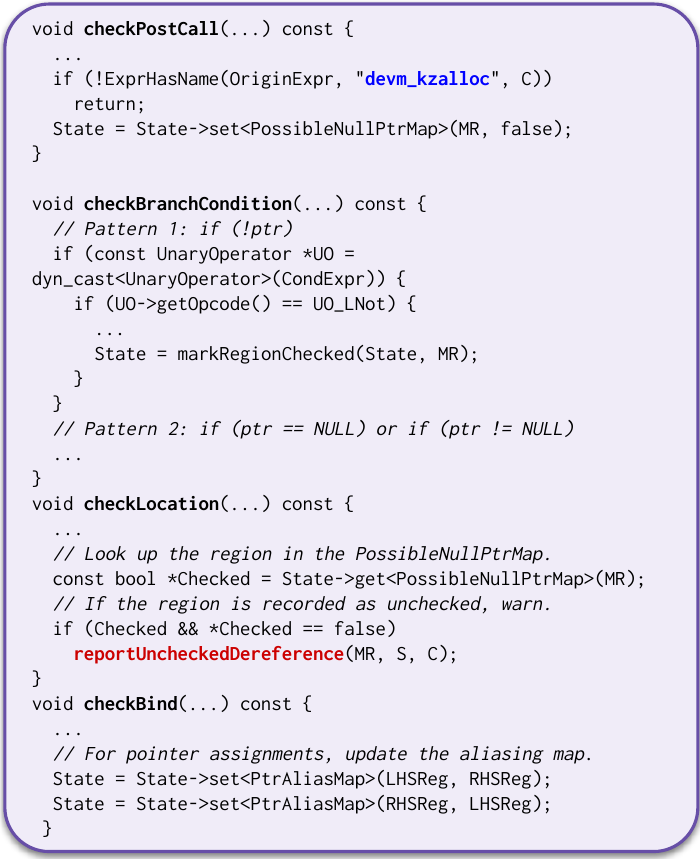}
\caption{A checker synthesized by \tech for the patch in Fig.~\ref{fig:kzalloc-patch}.}
\label{fig:kzalloc-checker}
\end{subfigure}

\caption{A bug pattern related to \texttt{devm\_kzalloc}.}
\end{figure}

\subsection{Clang Static Analyzer}

Static analysis~\cite{ayewah2008using} is a technique for detecting bugs by inspecting code without executing it. The {Clang Static Analyzer}~\cite{clang-static-analysis} (\csa) serves as a powerful engine for this purpose. It operates using path-sensitive symbolic execution, building an internal representation called an \CodeIn{ExplodedGraph}. Each node within this graph, an \CodeIn{ExplodedNode}, represents a specific \CodeIn{ProgramPoint} paired with an abstract \CodeIn{ProgramState}. This state meticulously maps program expressions to symbolic values and tracks the contents of memory locations.

The modularity of \csa is built upon checkers. These are small, specialized components, typically implemented as subclasses of a \CodeIn{Checker} template. Checkers function in an event-driven manner, registering interest in specific analysis events such as pre- and post-function calls, the identification of dead symbols, or instances of pointer escapes. A key capability of checkers is their ability to extend the \CodeIn{ProgramState} with custom, checker-specific data using provided macros, allowing them to maintain sophisticated state across the analysis.
Developing a new checker for \csa generally involves several steps: defining the specific bug pattern to be detected, implementing callback methods corresponding to the relevant analysis events, registering the new checker with the analysis framework, and integrating it into the testing system. Effective bug reporting is crucial, utilizing mechanisms like \CodeIn{BugType} and \CodeIn{BugReport} to provide clear diagnostics.

To illustrate, consider the example checker shown in Figure~\ref{fig:kzalloc-checker}, which registers four distinct callback functions.

\begin{itemize}
    \item The \CodeIn{checkPostCall} callback activates after function calls. It uses \CodeIn{ExprHasName} to check if the call was to \CodeIn{devm\_kzalloc}. If so, it updates the custom state map \CodeIn{PossibleNullPtrMap} to mark the returned memory region as potentially null (unchecked).
    
    \item The \CodeIn{checkBranchCondition} callback is used to handle conditional checks involving the pointer. It recognizes patterns like negation (\CodeIn{if (!ptr)}) or direct comparison (\CodeIn{if (ptr == NULL)}) and updates the state via \CodeIn{markRegionChecked} to reflect that a null check has occurred for the associated memory region.
    
    \item The \CodeIn{checkLocation} callback is triggered when a memory location is accessed. More specifically, it consults the \CodeIn{PossibleNullPtrMap} state; if the region is marked as unchecked at this point, it issues a warning using \CodeIn{reportUncheckedDereference}.
    
    \item The \CodeIn{checkBind} callback manages pointer assignments. It updates another custom state map, \CodeIn{PtrAliasMap}, to track potential aliases between memory regions involved in the assignment, ensuring the checker correctly handles cases where multiple pointers might refer to the same potentially null memory.
\end{itemize}

\begin{figure*}[t]
    \centering
    \includegraphics[width=0.96\textwidth]{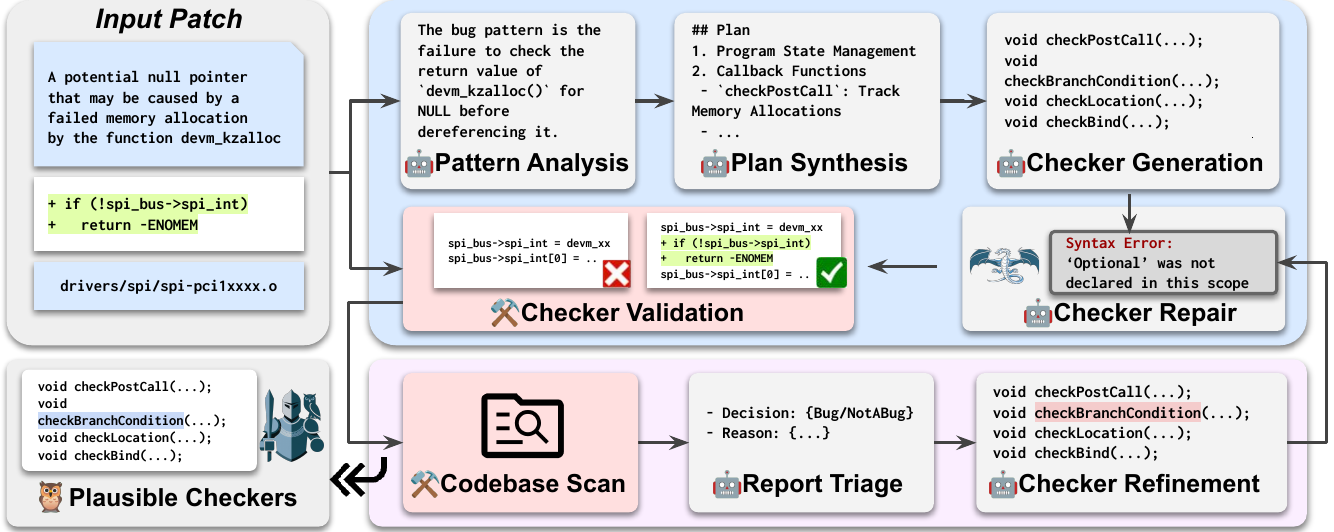}
    \caption{Overview of \tech.}
    \label{fig:overview}
\end{figure*}

\subsection{Motivating Example}

We demonstrate \tech's effectiveness through a case study involving a Null-Pointer-Dereference vulnerability pattern. Figure~\ref{fig:kzalloc-patch} shows a historical patch addressing this pattern, where the original bug stemmed from a missing null pointer check after a \CodeIn{devm\_kzalloc} call. Without this check, the system could crash if memory allocation failed and the returned null pointer was subsequently dereferenced.

\parabf{Limitations of existing tools.}
Despite this vulnerability pattern recurring since at least 2017 (commit \CodeIn{49af64e}), with our analysis identifying at least six historical patches addressing it, no static analysis tool had been developed to systematically detect these issues. Even specialized kernel checkers like \smatch~\cite{smatch} fail to identify these vulnerabilities because they lack the domain-specific knowledge that \CodeIn{devm\_kzalloc} may return \CodeIn{NULL} upon failure.

\parabf{Our approach.}
\tech extracts critical insights from the patch: unchecked return values from \CodeIn{devm\_kzalloc} represent potential Null-Pointer-Dereference vulnerabilities. The synthesized checker (written in \csa, Figure~\ref{fig:kzalloc-checker}) tracks null-check status across execution paths while correctly handling pointer aliasing, a sophisticated static analysis capability.
This checker discovered 3 \emph{new} vulnerabilities in the Linux kernel. Figure~\ref{fig:kzalloc-bug} presents one such vulnerability exhibiting the same pattern where a null pointer check is missing for the pointer returned by the \CodeIn{devm\_kzalloc} call.
This bug was subsequently fixed and assigned CVE-2024-50103.

\parabf{Advantages over direct \llm scanning.}
Directly using \llm{s} to scan the Linux kernel would be prohibitively expensive, as \CodeIn{devm\_kzalloc} alone appears over 7K times across 5.4K files. In contrast, \tech's static analyzers primarily consume CPU resources rather than repeated \llm invocations, making the approach both scalable and cost-effective.
Moreover, since generating the checkers is mostly a one-time effort, they can naturally evolve alongside the system.

\parabf{Technical challenges and solutions.}
Creating effective static analyzers with \llm{s} presents several challenges. First, writing robust checkers end-to-end is complex. \tech addresses this through a multi-stage synthesis pipeline that breaks down complex tasks into manageable steps. Second, \llm hallucination can produce incorrect analyzers. \tech mitigates this by validating synthesized checkers against historical patches, verifying they correctly distinguish between buggy and patched code. Finally, to reduce false positives, we implement a bug triage agent that identifies false alarms, enabling iterative refinement of the checkers.

\section{Design}

\parabf{Terminology.}
\tech takes a patch commit as input and outputs a corresponding \csa checker. \emph{Valid checkers} correctly distinguish between buggy and patched code, flagging pre-patch code as defective while recognizing post-patch code as correct. \emph{Plausible checkers}\footnote{We adopt the term ``plausible'' from program repair~\cite{qi2015analysis, xia2023automated}, where a ``plausible'' patch passes all test cases and potentially is the correct fix.} are \emph{valid checkers} that additionally demonstrate practical utility through low false positive rates or a manageable number of reports. We provide formal definitions of these terms in \S~\ref{sec:impl}.

\parabf{Overview.}  
\tech leverages agentic workflow to process patch commits for static analyzer synthesis, as illustrated in Figure~\ref{fig:overview}.  
It operates in two phases: checker synthesis (\S~\ref{subsec:checker-syn}) and checker refinement (\S~\ref{subsec:checker-evo}).  
In the checker synthesis phase, \tech analyzes the input patch to identify bug patterns (\S~\ref{subsec:pattern}), synthesizes a detection plan (\S~\ref{subsec:plan}), and implements a checker using \csa (\S~\ref{subsec:checker-impl}).
If compilation errors occur, a syntax-repair agent automatically repairs them based on the error messages.  
This phase concludes with the generation of \emph{valid checkers} (\S~\ref{subsec:validation}).
In the subsequent checker refinement phase, these valid checkers are deployed to scan the entire codebase for potential bugs.  
When bug reports are generated, a triage agent evaluates them for false positives, and \tech refines the checker accordingly.  
If the scan produces a manageable number of reports with a low false positive rate, \tech presents the \emph{plausible checkers} and their filtered reports as potential bugs for review.

\begin{figure}[t]
    \centering

\centering
\includegraphics[width=0.95\linewidth]{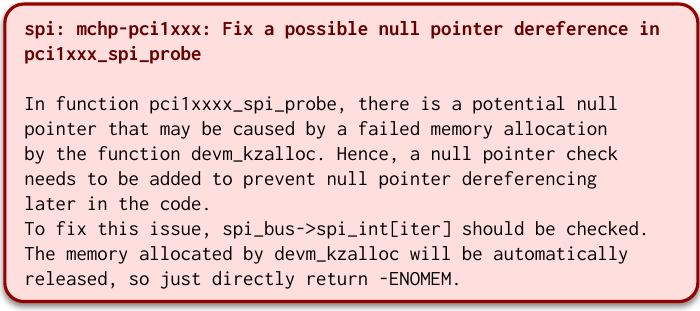}
    
    \caption{Patch commit message.}
    \label{fig:alloc-patch-msg}
\end{figure}

\subsection{Checker Synthesis}
\label{subsec:checker-syn}

Algorithm~\ref{algo:sagen} presents the multi-stage pipeline of checker synthesis.  
In the first stage, \tech analyzes the bug pattern shown in the patch (\CodeIn{Line~\ref{line:analyze}}).  
Next, \tech synthesizes the plan based on the patch and the identified bug pattern (\CodeIn{Line~\ref{line:plan}}).  
With the plan in hand, \tech implements the checker using \csa (\CodeIn{Line~\ref{line:impl}}).  
If any compilation issues arise, a syntax-repair agent is invoked to debug and repair them (\CodeIn{Line~\ref{line:repair}}).  
The repair process is allowed up to \CodeIn{maxAttempts} (default is 5) attempts.  
If the checker compiles successfully, \tech validates it by checking whether it can distinguish between the buggy and patched code (\CodeIn{Line~\ref{line:valid}}).  
Once the checker is deemed valid, it is returned for the next phase (\CodeIn{Line~\ref{line:return-valid}}).  
Otherwise, the synthesis pipeline continues iterating until reaching \CodeIn{maxIterations}.  
If all iterations fail, the process returns Null, indicating that a valid checker could not be synthesized (\CodeIn{Line~\ref{line:return-null}}).

\begin{algorithm}[t]
\small
\caption{Synthesize checkers with input patch.}
\label{algo:sagen}
\DontPrintSemicolon
\SetKwProg{Fn}{Function}{:}{}
\SetKwFunction{ProcessPatch}{GenChecker}
\SetKwFunction{GenerateDemo}{GenerateDemo}
\SetKwFunction{SynthesizeChecker}{SynthesizeChecker}
\SetKwFunction{EvaluateChecker}{ValidateChecker}

\SetKwFunction{AnalyzePatch}{AnalyzePatch}
\SetKwFunction{SynthesizePlan}{SynthesizePlan}
\SetKwFunction{RefineWithExpert}{RefineWithExpert}
\SetKwFunction{ImplementChecker}{Implement}
\SetKwFunction{RepairChecker}{RepairChecker}
\SetKwFunction{GenerateTests}{GenerateTests}

\Fn{\ProcessPatch{patch}}{
    \textit{\color{purple} \# Iterative checker generation and evaluation\;}
    \For{i = 1 \KwTo maxIterations}{
    \textit{\color{purple} \# Stage 1: Bug Pattern Analysis\;}
    pattern $\leftarrow$ \AnalyzePatch{patch}\;\label{line:analyze}
    
    \textit{\color{purple} \# Stage 2: Detection Plan Synthesis\;}
    plan $\leftarrow$ \SynthesizePlan{patch, pattern}\;\label{line:plan}
    
        \textit{\color{purple} \# Stage 3: Analyzer Implementation and Repair\;}
        checker $\leftarrow$ \ImplementChecker{patch, pattern, plan}\;\label{line:impl}
        attempts $\leftarrow$ 0\;
        \While{hasCompilationErrors(checker) \textsc{and} attempts < maxAttempts}{
            checker $\leftarrow$ \RepairChecker{checker}\; \label{line:repair}
            attempts $\leftarrow$ attempts + 1\; 
        }
        
        \If{hasCompilationErrors(checker)}{
            \textit{\color{purple} \# Skip evaluation if checker still has errors\;}
            Continue\;  \label{line:repair-fair}
        }
        
        \textit{\color{purple} \# Stage 4: Validation\;}
        isValid $\leftarrow$ \EvaluateChecker{checker, patch}\; \label{line:valid}
        
        \If{isValid}{
            \Return checker\; \label{line:return-valid}
        }
    }
    \Return Null\; \label{line:return-null}
}
\end{algorithm}

\begin{figure}[t]
    \centering

\begin{subfigure}[t]{\linewidth}
\centering
\includegraphics[width=0.9\linewidth]{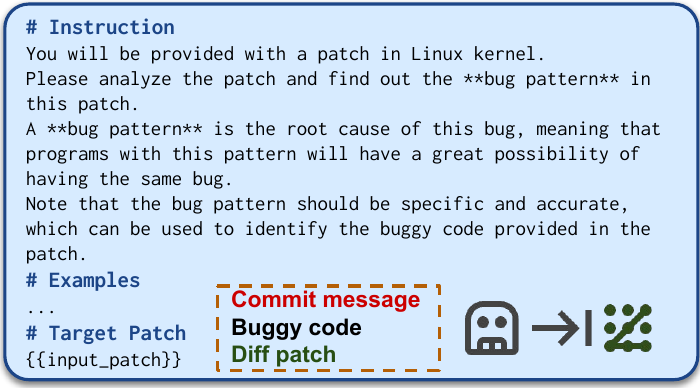}
    \caption{Prompt template for bug pattern analysis}
    \label{fig:analyze-template}
\end{subfigure}

\begin{subfigure}[t]{\linewidth}
\centering
\includegraphics[width=0.9\linewidth]{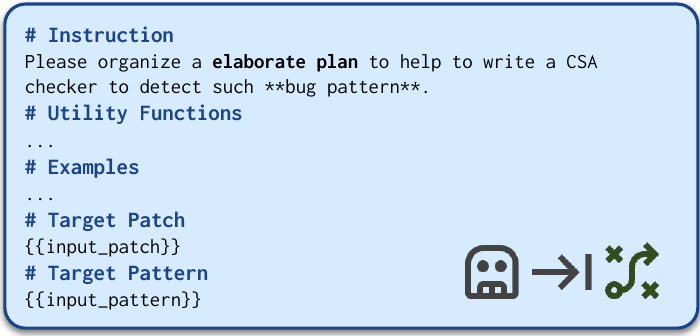}
    \caption{Prompt template for plan synthesis.}
    \label{fig:plan-template}
\end{subfigure}

\begin{subfigure}[t]{\linewidth}
    \centering
    \includegraphics[width=0.9\linewidth]{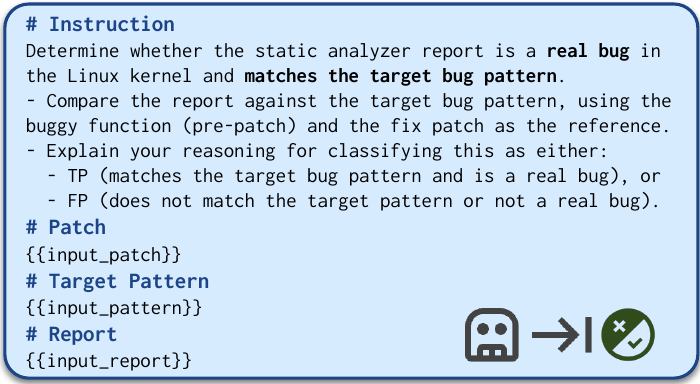}
    \caption{\revision{Prompt template for report triage.}}
    \label{fig:triage-template}
\end{subfigure}

    \caption{\revision{Simplified prompt templates used by \tech.}}
    \label{fig:template}
\end{figure}

\subsubsection{Bug Pattern Analysis}
\label{subsec:pattern}

The initial stage involves analyzing patch commits to identify underlying bug patterns.
Patch commits typically consist of \CodeIn{diff} patches and may include developer comments describing the bug being fixed, as illustrated in Figure~\ref{fig:alloc-patch-msg}.
Our goal is to extract patterns that can be translated into static analysis rules for bug detection.
While bug patterns are sometimes explicitly described in commit messages, they often require deeper analysis of the code changes within the patch.

We have developed an \llm-based agent specifically designed to perform this pattern analysis, with the prompt template shown in Figure~\ref{fig:analyze-template}. In addition to the patch, we extract the complete function code that was modified from the kernel codebase.
This additional context is crucial because the patch \CodeIn{diff} alone may not capture all relevant buggy patterns, as some issues depend on the broader context of the code. By providing both the patch and the complete function code to \llm{s}, we enable a more comprehensive understanding of the bug being patched.

A single bug pattern identified from a patch can be expressed with varying scope and complexity. Consider the Null-Pointer-Dereference involving \CodeIn{devm\_kzalloc} (Figure~\ref{fig:kzalloc-patch}). A \textit{broad} pattern (e.g., check \textit{any} potentially null return) is comprehensive, but identifying all relevant functions/conditions poses significant static analysis challenges, hindering robust implementation by \llm{s}. Consequently, our approach favors more \textit{targeted} bug patterns derived from the patch context. These facilitate precise and tractable checker synthesis by the \llm{s}. For the \CodeIn{devm\_kzalloc} example, focusing specifically on its return value yields a targeted pattern that effectively addresses the observed bug class while being significantly more manageable for the \llm to implement correctly compared to the broader, more complex alternative.

\subsubsection{Plan Synthesis}
\label{subsec:plan}

Once the bug pattern is identified, \tech generates a high-level plan for implementing the static analyzer.
This plan serves two critical purposes: first, it provides structured guidance to the \llm{s} during implementation, preventing confusion and promoting effective execution.
Second, it facilitates debugging of the entire pipeline by making the \llm{s}' reasoning process transparent and traceable.
Our ablation study in \S~\ref{subsec:ablation} confirms the value of this plan synthesis, demonstrating improved performance consistent with findings in other domains~\cite{sun2023pearl}.

For instance, synthesizing a checker for the unchecked \CodeIn{devm\_kzalloc} return value pattern (illustrated in Figure~\ref{fig:kzalloc-checker}) might generate a plan with key steps such as: (1) Using program state to track memory regions from \CodeIn{devm\_kzalloc}, (2) monitoring conditional branches (\CodeIn{checkBranchCondition}) to mark regions as checked if a null check occurs, and (3) detecting uses (\CodeIn{checkLocation}) of unchecked regions, potentially signaling a bug. This high-level structure guides the subsequent implementation phase.

To synthesize the implementation plan for the checker, we have designed an \llm-based agent whose prompt template is shown in Figure~\ref{fig:plan-template}. This agent takes the previously summarized bug pattern as input. Additionally, we maintain a curated database of utility functions for checker implementation that can be easily extended. By including the signatures and brief descriptions of these utility functions in the prompt, we enable \llm{s} to leverage them effectively during the planning process, simplifying the overall task.

\subsubsection{Analyzer Implementation and Syntax Repair}
\label{subsec:checker-impl}

\begin{figure}
    \centering
    \includegraphics[width=0.9\linewidth]{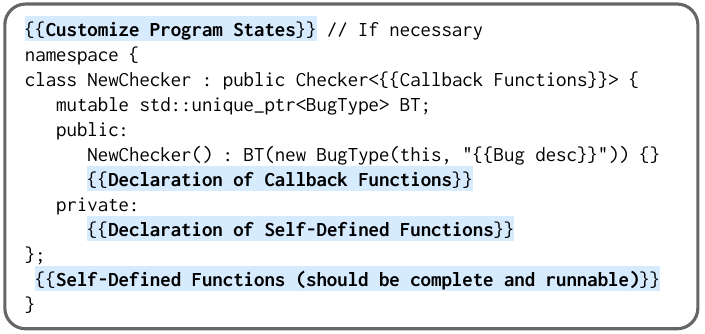}
    \caption{Pre-defined checker template for \csa.}
    \label{fig:checker-template}
\end{figure}
After identifying the bug pattern and making the plan, we leverage an \llm-based agent to implement the corresponding checker.
To maximize implementation accuracy, we provide the agent with comprehensive inputs: the distilled bug pattern and a structured implementation plan.
We also provide a pre-defined checker template, as shown in Figure~\ref{fig:checker-template}, which standardizes the implementation structure and reduces potential errors.
Moreover, we provide a list of utility functions that could help with the implementation.

The synthesized checkers could have compilation errors, \eg using the wrong static analysis API or incorrect variable types.
To handle the potential compilation errors, we employ a dedicated debugging agent.
Inspired by existing work on program repair~\cite{xia2024automated}, this agent automatically processes compiler error messages and applies necessary fixes, effectively addressing syntax errors that may arise from \llm hallucinations.
This automated debugging pipeline ensures that the final checkers are both syntactically correct and compilable.

\subsubsection{Validation}
\label{subsec:validation}

To semantically validate our checkers and mitigate potential inaccuracies from \llm{s}, we evaluate them against both the buggy (pre-patch) and patched versions of the relevant code (e.g., Linux kernel files).
This differential analysis verifies that a checker correctly identifies the target bug in the original code and confirms its absence after the patch.
For efficiency, we scope this validation to only the files modified by the patch and their dependencies, rather than the entire codebase.
A checker is considered \emph{valid} if it flags the bug in the pre-patch version and shows a corresponding reduction or elimination of that specific warning in the patched version. More details are in \S~\ref{sec:impl}.

\subsection{Checker Refinement}
\label{subsec:checker-evo}

Following synthesis, each \emph{valid} checker is used to scan the entire system. However, its initial validation doesn't prevent potential false positives when analyzing the broader codebase, where \emph{correct} code might be flagged erroneously. To mitigate this, we implement an iterative refinement procedure driven by \llm{s}. This involves evaluating the generated bug reports and feeding identified false positives back to refine the checker.

However, automating this refinement faces hurdles. First, bug reports are often verbose, containing extensive context that is difficult to process efficiently. Second, debugging the checker logic and modifying it correctly based on false positives requires complex analysis.

Our refinement pipeline addresses these challenges methodically. First, to manage report complexity, we distill generated bug reports to their essential components---primarily the ``relevant lines'' highlighted by the static analyzer (\eg{} \csa~\cite{clang-static-analysis}) and the corresponding trace path—stripping extraneous context while preserving critical diagnostics.
\revision{Second, to navigate the complexity of analysis and modification, we employ specialized \llm{}-based agents. A triage agent classifies each distilled report, focusing strictly on alignment with the target bug pattern (rather than general code correctness); the prompt template is shown in Figure~\ref{fig:triage-template}.}

\begin{figure}
    \centering
    \includegraphics[width=0.9\linewidth]{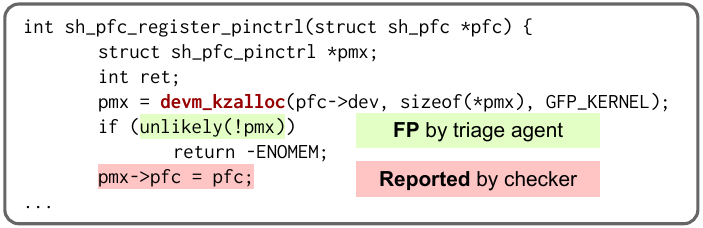}
    \caption{A report labeled as FP by our triage agent.}
    \label{fig:triage-example}
\end{figure}

If the triage agent identifies a report as a false positive, as exemplified in Figure~\ref{fig:triage-example}, a dedicated refinement agent takes over. In the case shown, the initial checker (derived from the patch in Figure~\ref{fig:kzalloc-patch}) flagged the use of \CodeIn{pmx->pfc} because its logic failed to recognize \CodeIn{if (unlikely(!pmx))} as a valid null check, perhaps confused by the \CodeIn{unlikely()} macro.
The triage agent correctly interprets the check semantically and flags the report as FP.
The refinement agent then uses this information to adjust the checker's logic, specifically enhancing its ability to handle constructs like \CodeIn{unlikely()}, thereby preventing this type of false positive in subsequent scans while ensuring it can still detect the original vulnerability.

A refined checker is accepted only if it satisfies two criteria: (1) it no longer generates warnings for the previously identified false positive cases, and (2) it maintains its validity by correctly differentiating between the original buggy and patched code versions.
This criterion ensures the semantic accuracy of the refined checkers.

\section{Implementation}
\label{sec:impl}
\label{sec:impl}

\parabf{Input commit collection.}
To collect patch commits for rigorous evaluation, we implemented a systematic classification and selection process. First, we established 10 distinct bug categories. We then used relevant keywords to identify potentially related commits. A commit was included in our dataset only when two authors independently agreed on its categorization.
For each bug type, we initially examined the first 20 commits that matched our search criteria. We continued reviewing commits beyond the initial 20 if we hadn't yet collected 5 qualifying commits for a given category. Our goal was to gather a minimum of 5 commits per bug type whenever possible.
Table~\ref{tab:collect-commits} presents our categorization of 10 bug types and their corresponding patch commit counts. These manually collected and labeled commits served as our benchmark dataset for rigorous evaluation.

\begin{table}[t]\centering
\caption{Distribution of \textbf{patch commits} across 10 bug categories and the validity status of their synthesized checkers. ``NPD'' denotes ``Null-Pointer-Dereference'' and ``UBI'' indicates ``Use-Before-Initialization''.}
\label{tab:collect-commits}
\small
\begin{tabular}{l|r|r|rrrr}\toprule
& & &\multicolumn{3}{c}{Valid} \\\cmidrule{4-6}
Bug Type &Total &Invalid &Direct &Refined &Fail \\\midrule
NPD &6 &1 &2 &2 &1 \\
Integer-Overflow &7 &3 &1 &3 &0 \\
Out-of-Bound &6 &2 &4 &0 &0 \\
Buffer-Overflow &5 &3 &2 &0 &0 \\
Memory-Leak &5 &2 &3 &0 &0 \\
Use-After-Free &7 &4 &2 &1 &0 \\
Double-Free &8 &1 &5 &1 &1 \\
UBI &5 &1 &1 &3 &0 \\
Concurrency &5 &2 &3 &0 &0 \\
Misuse &7 &3 &3 &1 &0 \\ \midrule
Total &61 &22 &26 &11 &2 \\
\bottomrule
\end{tabular}
\end{table}

\parabf{Few-shot examples.}
We prepared three end-to-end examples for in-context learning.
These three are patch commit \CodeIn{3027e7b15b02} (Null-Pointer-Dereference), \CodeIn{3948abaa4e2b} (Use-Before-Initialization), and \CodeIn{4575962aeed6} (Double-Free).
The design and implementation of the checker for these three commits required approximately 40 person-hours.
This was a one-time effort, yielding reusable examples.
We also explore the use of real-world, off-the-shelf examples (\S~\ref{subsec:ablation}).

\parabf{Utility functions.}
While implementing example checkers, we identified several common helper operations. We implemented 9 such utility functions (e.g., \CodeIn{getMemRegionFromExpr}) to encapsulate low-level Clang Static Analyzer tasks, simplifying checker development, particularly for \llm synthesis. These utilities were designed for simplicity and extensibility.

\parabf{Valid checkers.}
To evaluate checker validity, we verify that it can both detect the original bug and recognize its fix.
We first identify \emph{buggy objects} by examining the modified files in the diff patch.
Next, we check out the repository to the buggy commit (immediately preceding the patch) and scan these objects to count the number of bug reports ($N_{buggy}$).
We then scan these objects after applying the patch commit to obtain the number of remaining bug reports ($N_{patched}$).
A checker is considered valid if $N_{buggy} > N_{patched}$ and $N_{patched} < T_{valid}$, where $T_{valid}$ is a threshold value (50 by default).

\parabf{Plausible checkers.}
We determine \correct checkers based on their performance when analyzing the entire Linux kernel.
Our approach is founded on the principle that high-quality checkers, especially those derived from historical commits, should generate a reasonable number of actionable bug reports.
A checker is classified as \correct if it either: (1) produces fewer reports than a predefined threshold $T_{\correct}$ (default: 20), or (2) demonstrates an acceptable false positive rate in sampled warnings.

\parabf{Checker refinement.}
We evaluate each valid checker by scanning the entire kernel codebase independently, with execution bounded by either a one-hour time limit or a maximum of 100 warnings during the refinement process.
Note that these limits are only applied during the checker refinement phase; when performing actual bug detection, we run the checkers without such constraints.
The refinement process begins with \llm{}-assisted triage of the checker's output.
Using a consistent random seed, we sample 5 warnings for \llm{} inspection due to cost consideration.
A checker qualifies as \correct if it either generates fewer than $T_{\correct}=20$ total reports or exhibits at most one false positive in the evaluated sample (labeled by our triage agent).
For checkers failing these criteria, we implement an iterative refinement protocol targeting the identified false positives, permitting up to three refinement iterations to improve precision.

\section{Evaluation}

We explore the following research questions for \tech:

\begin{enumerate}[label=\textbf{ RQ-\arabic{enumi}.},leftmargin=*]
    \item Can \tech generate high-quality checkers?
    \item Can the checkers generated by \tech find real-world kernel bugs?
    \item Are the capabilities of \tech orthogonal to the human-written checkers?
    \item Are all the key components in \tech effective?
\end{enumerate}

\parabf{Evaluation metrics.}
We conduct an extensive evaluation by using the following metrics: 

\textit{Checker Validity Rate.}  
A valid checker successfully identifies the buggy pattern in the original code and confirms its absence in the patched version. This metric reflects our framework's and \llm{s}' ability to understand patch semantics and synthesize discriminative checkers.

\textit{Plausible Checker Rate.}  
This metric measures the number of high-quality checkers synthesized, representing those that are both valid and exhibit a low false positive rate.

\textit{Bug Detection.}  
We assess the number of real-world bugs successfully detected by the synthesized checkers.

\textit{Resource Efficiency.}  
This metric captures the computational time and monetary costs associated with both checker synthesis and execution.

\textit{Checker Error Categories.}  
We classify checker failures into the following categories, ordered by severity:
\begin{itemize}[topsep=1pt]
    \item {Compilation Failures:} Checkers that fail during compilation due to syntax or dependency errors.
    \item {Runtime Errors:} Checkers that compile successfully but crash during execution (e.g., \textit{"The analyzer encountered problems on source files"}).
    \item {Semantic Issues:} Checkers that cannot distinguish between the buggy and patched code.
\end{itemize}

\textit{Static Analysis Capabilities.}
We further examine the static analysis capabilities employed by checkers, including path sensitivity, region sensitivity, and advanced state tracking.

\parabf{Hardware and software.}  
All our experiments are run on a workstation with 64 cores, 256 GB RAM, and 4 Nvidia A6000 GPUs, operating on Ubuntu 20.04.5 LTS.  
We use O3-mini as our default \llm backend.
By default, when scanning the entire codebase, we use \CodeIn{-j32}.  
We evaluated using Linux v6.13, and for bug finding, we examined versions from v6.9 to v6.15.  
The Linux configuration used is \CodeIn{allyesconfig}.

\subsection{RQ1: Synthesized Checkers}

\begin{figure}[t]

\begin{subfigure}[t]{0.95\linewidth}
\includegraphics[width=\linewidth]{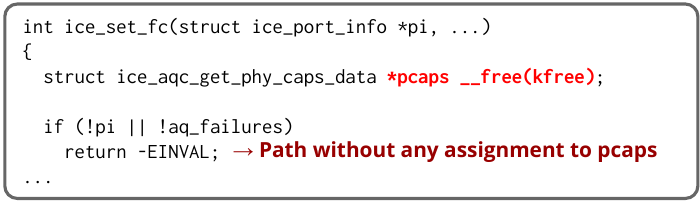}
\caption{Bug in Use-Before-Initialization patch.}
\label{fig:uninit-ptr-patch}
\end{subfigure}

\begin{subfigure}[t]{0.95\linewidth}
\includegraphics[width=\linewidth]{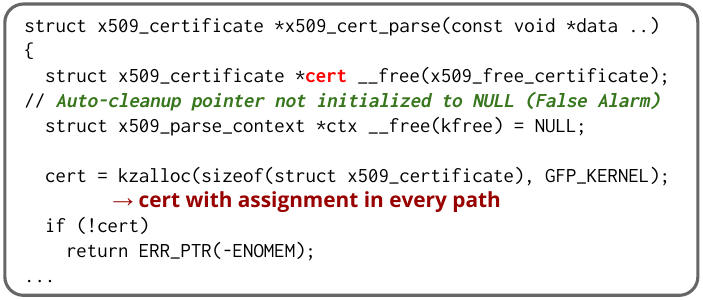}
\caption{False positive report for UBI.}
\label{fig:uninit-ptr-nbug}
\end{subfigure}

\caption{Examples of false positives by \tech.}
\end{figure}
We evaluate \tech on the 61 commits listed in Table~\ref{tab:collect-commits} to show that it can synthesize high-quality checkers across various bug types, beyond those in our few-shot examples.

\subsubsection{Checker Synthesis}
In total, valid checkers were generated for \numValid commits. The 39 synthesized checkers average 125.7 lines of code and exhibit diverse static analysis capabilities. Specifically, 37 checkers are path-sensitive, 13 incorporate region sensitivity, and 16 employ advanced state tracking; in contrast, only 2 leverage AST travelers.
This suggests that \tech can generate complex analysis logic, not just simple pattern matching.

\parabf{Cost.}
The full synthesis process required 15.9 hours, during which 8.2 million input tokens were processed and 1.2 million output tokens produced, resulting in an approximate cost of \emph{\$0.24 per commit} using O3-mini. For commits that ultimately yielded valid checkers, an average of 2.4 synthesis attempts was necessary (with a maximum of 8 attempts observed).

\parabf{Failure analysis.}
We now break down the failures from two perspectives: the underlying \emph{failure root causes} and the observed \emph{failure symptoms} during the synthesis process.

\emph{(i) Failure root causes}. Among the 61 commits processed, 22 did not result in any valid checker. Our investigation into these failures indicates that:
\begin{itemize}[topsep=1pt]
    \item 2 (9\%) commits failed due to an inaccurate bug pattern,
    \item 7 (32\%) failed owing to an inaccurate plan, and
    \item 13 (59\%) were caused by inaccurate implementation.
\end{itemize}
For the implementation-related failures, a common issue was that compiler optimizations inlined certain function calls (e.g., \CodeIn{strcp} and \CodeIn{memset}), which prevented the checker from properly intercepting these calls.

Our approach exhibits limitations in handling buffer overflow and use-after-free commits. We believe these challenges stem from two main factors: static analysis inherently struggles with precise value determination---especially when establishing buffer bounds at compile time---and it also faces significant hurdles in analyzing multi-threaded code, for instance, when assessing the proper use of locks.

\emph{(ii) Failure symptoms.}
During synthesis (allowing up to 10 attempts per commit), 273 failed attempts were recorded from all 61 commits. The failures can be categorized as:
\begin{itemize}[topsep=1pt]
    \item 65 attempts (23.8\%) resulted in compilation errors,
    \item 1 attempt (0.4\%) led to a runtime error, and
    \item 207 attempts (75.8\%) suffered from semantic issues that obstructed proper bug identification.
\end{itemize}
Of the 207 semantic failures, 34 checkers erroneously flagged both buggy and patched code as potentially problematic, while the remaining 173 misclassified both versions as bug-free. This outcome underscores the challenge of accurately distinguishing buggy code.

Interestingly, even the 173 checkers that did not recognize the specific bug in their input patch can still be valuable. When deployed across the large system, these checkers may successfully detect bugs with similar patterns in other contexts. This apparent paradox likely arises because the failure to detect the training bug is sometimes due to edge cases or context-specific complexities rather than inherent deficiencies in the checkers' detection logic. Moreover, these checkers generally exhibit lower false positive rates compared to those that incorrectly flag both buggy and patched code, enhancing their practical utility for bug detection.

\subsubsection{Checker Refinement.}
After scanning the entire kernel codebase with these \numValid valid checkers, 26 of them were labeled ``\correct'' directly.
Our refinement pipeline was applied to the remaining 13 valid checkers, successfully refining 11 of them.
In total, 19 refinement steps were completed successfully.
This demonstrates the effectiveness of our refinement pipeline, which successfully refined 84.6\% of the valid checkers that were not ``\correct'' initially.

\parabf{False positive rate.}
Of the 37 \correct checkers, 16 did not report any bugs.
For the remaining checkers, we applied our bug triage agent to filter all the reports, focusing only on those labeled as ``bug'' since our triage agent demonstrated a low false negative rate in our evaluation (as shown in \S~\ref{subsec:triage}).
In total, we obtained 90 reports labeled as ``bug''. Upon manual verification, we confirmed 61 true positives.
This indicates that the combination of our \correct checkers and bug triage agent has a false positive rate of 32.2\%.

Our manual analysis of the 29 false positives revealed three recurring patterns leading to incorrect reports:
\begin{itemize}[topsep=1pt]
    \item \emph{Inaccurate bug pattern:} In 5 cases, although the checker correctly identified the original bug/patch scenario, the inferred bug pattern lacked the necessary precision for reliable detection across different contexts.
    \item \emph{Incorrect pattern matching:} For 6 reports, the checker correctly identified the bug pattern but applied it too broadly, flagging code segments that did not meet the specific constraints intended by the pattern.
    \item \emph{Trigger condition mismanagement:} The most common issue (18 reports) involved checkers where both the pattern and matching were correct, but the checker failed to manage trigger or state conditions properly (e.g., failing to recognize a pointer had already been validated before use).
\end{itemize}

\emph{Case study: A high false positive rate checker.}  
In commit \CodeIn{90ca6956d383} (``ice: Fix freeing uninitialized pointers'', see Figure~\ref{fig:uninit-ptr-patch}), the issue stems from the pointer \CodeIn{pcaps} not being initialized to \CodeIn{NULL}. If an early return or error path occurs before the pointer is allocated, the cleanup routine may inadvertently attempt to free an uninitialized (or garbage) pointer. In such cases, the checker should account for the possibility of an early exit leaving the pointer unset.
In contrast, the bug report in Figure~\ref{fig:uninit-ptr-nbug} highlights a scenario where, although the \CodeIn{cert} pointer starts uninitialized, it is immediately assigned a valid value along every execution path, ensuring it is never left in an unassigned state. Thus, despite the initial uninitialized state, the code does not constitute a bug.
Our synthesized checker did not incorporate these nuanced constraints, and the triage agent likewise failed to recognize the critical differences, ultimately leading to a false positive.

\begin{table}[]
    \centering
    \small
        \caption{Newly detected bugs by \tech.}
        \begin{tabular}{l|r|rrr|r}
        \toprule
                   & Total & Confirmed & Fixed & Pending & CVE \\
        \midrule
        \tech      & \numBug & \numConfirmed & \numFixed & \numPending & \numCVE \\
        \bottomrule
        \end{tabular}
    \label{tab:bug}
\end{table}

\begin{figure}[h!]
    \centering %
    \begin{subfigure}[b]{0.43\textwidth}
        \centering
        \includegraphics[width=0.95\linewidth]{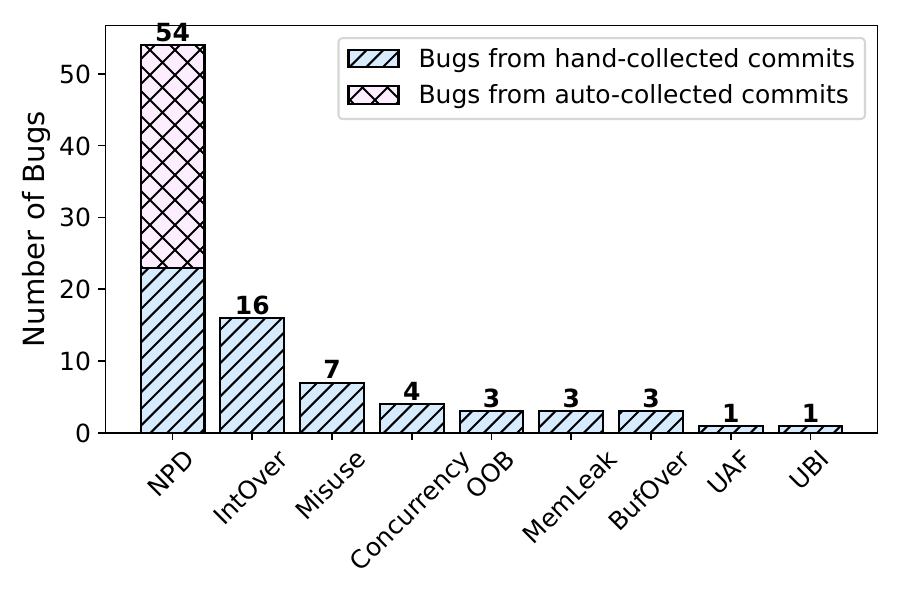}
        \caption{Number of bugs in each type.}
        \label{fig:bug-type}
    \end{subfigure}

    \begin{subfigure}[b]{0.43\textwidth}
        \centering
        \includegraphics[width=0.95\linewidth]{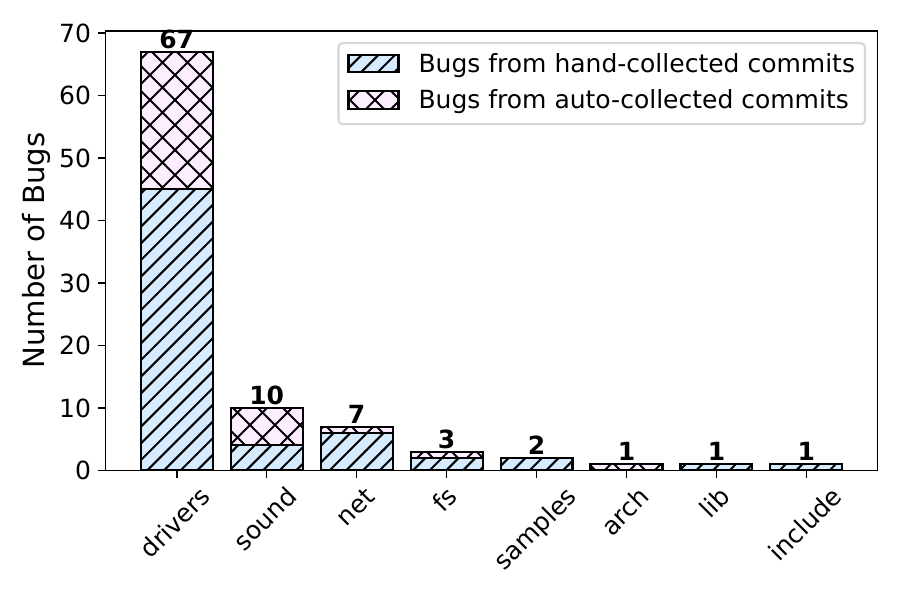}
        \caption{Number of bugs in each subsystem.}
        \label{fig:bug-location}
    \end{subfigure}
    \hfill %
    \begin{subfigure}[b]{0.43\textwidth}
        \centering
        \includegraphics[width=0.95\linewidth]{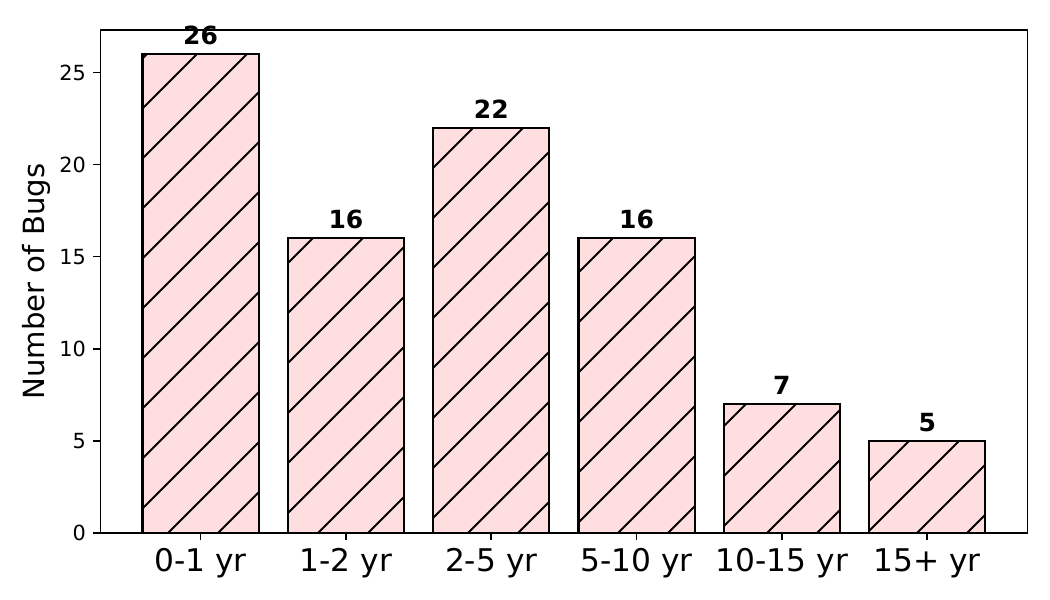}
        \caption{Number of bugs with different lifetimes.}
        \label{fig:bug-time}
    \end{subfigure}
    \hfill
    \begin{subfigure}[b]{0.43\textwidth}
        \centering
        \includegraphics[width=0.95\linewidth]{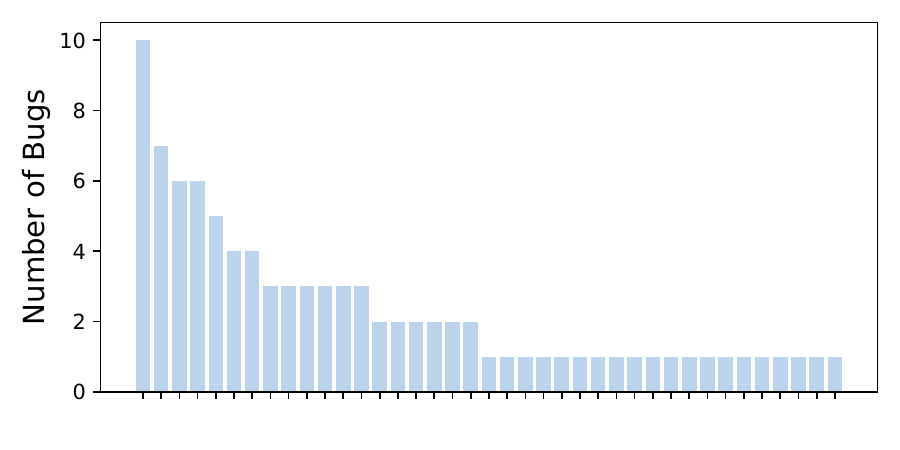}
        \caption{\revision{Number of bugs detected by each commit.}}
        \label{fig:bug-commit}
    \end{subfigure}

    \caption{Details of new bugs. Subfigures (\subref{fig:bug-type}), (\subref{fig:bug-location}), (\subref{fig:bug-time}), and  (\subref{fig:bug-commit}) show breakdowns by type, subsystem, lifetime, and commit.}
    \label{fig:bug-detail} %
\end{figure}

\subsection{RQ2: Detected Bugs}
\subsubsection{Overall.}  
To date, static analyzers synthesized by \tech have identified \numBug new bugs in the Linux kernel. As summarized in Table~\ref{tab:bug}, developer confirmation has been received for \numConfirmed of these bugs. Among those confirmed, \numFixed have already been fixed. The remaining \numPending bugs are currently awaiting developer review (calculated as Total - Confirmed). Notably, \numCVE of the discovered bugs have been assigned CVE numbers, showing their practical security impact.

The checkers for bug detection originate from two sources: (i) the initial 61 manually collected commits used for evaluation in \S~\ref{subsec:checker-syn} across diverse bug types (shown in Table~\ref{tab:collect-commits}), and (ii) an additional set of 100 commits automatically collected using keywords related to Null-Pointer-Dereference to further explore that specific bug class.
Figure~\ref{fig:bug-type} and Figure~\ref{fig:bug-location} show the bug distribution from these sources---with the \colorbox{evalColor}{light blue} for bugs from the initial evaluation set and the \colorbox{npdColor}{light purple} for those from the auto-collected set.

\parabf{Bug types.}  
Analysis of checkers from the manually collected commits reveals that \tech can detect a diverse range of bug types (see Figure~\ref{fig:bug-type}). Null-Pointer-Dereference bugs are the most prevalent, highlighting \tech’s strength in this area. In response, we expanded our effort by automatically collecting commits related to Null-Pointer-Dereference, which yielded an additional \numNewBug bugs.

\parabf{Bug location.}  
As shown in Figure~\ref{fig:bug-location}, the detected bugs span various Linux kernel subsystems. The majority appear in the \CodeIn{drivers} subsystem (67 out of \numBug), reflecting its large footprint in the kernel~\cite{bursey2024syzretrospector}. Additionally, 10 and 7 bugs were identified in the \CodeIn{sound} and \CodeIn{net} subsystems, respectively. Notably, 2 bugs were found in the \CodeIn{samples} directory---an area that provides example usage for kernel developers and where correctness is especially critical.

\parabf{Bug lifetime.}  
Figure~\ref{fig:bug-time} illustrates the distribution of bug lifetimes. Notably, the average bug lifetime is \emph{\avgYear years}, and \overfive bugs had existed for \emph{over five years} before detection.
This indicates that the vulnerabilities uncovered by \tech are difficult to detect and remain latent for extended periods, underscoring the effectiveness of our approach.

\parabf{\revision{Bug distribution over commits.}}
\revision{Synthesized checkers from 39 bug-fix commits uncovered new bugs (2.4 each on average), with a long-tail skew (as shown in Figure~\ref{fig:bug-commit}): five checkers each found five or more bugs. In general, checkers derived from recurring error patterns yield higher counts, while those from specialized fixes surface fewer, yet still valuable, findings. This suggests that \tech can learn and propagate impactful patterns beyond their original contexts, producing a mix of broad-coverage and high-yield checkers.}

\subsubsection{Case Study.}
Here are examples of vulnerabilities detected by \tech.

\begin{figure}[t!]
\centering
\begin{subfigure}[t]{\linewidth}

\centering
\includegraphics[width=\linewidth]{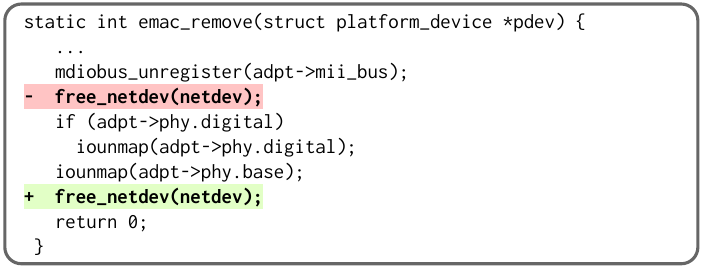}
\caption{{Input Use-After-Free patch.} }
\label{fig:uaf-patch}
\end{subfigure}

\begin{subfigure}[t]{\linewidth}
\centering
\includegraphics[width=\linewidth]{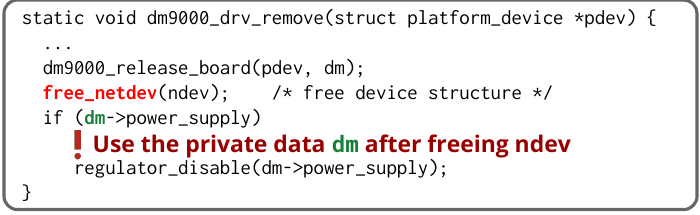}
\caption{\textbf{CVE-2025-21715}, found by the checker for the patch above.}
\label{fig:uaf-bug}
\end{subfigure}

\begin{subfigure}[t!]{\linewidth}

\centering
\includegraphics[width=\linewidth]{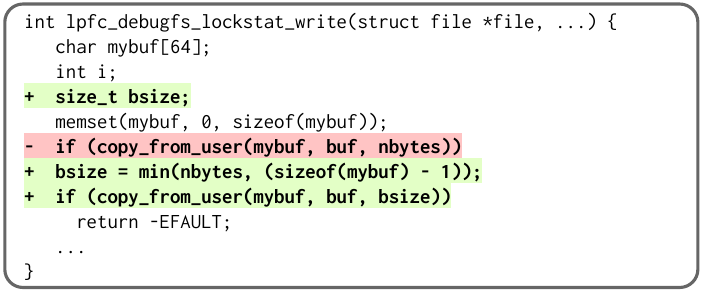}
\caption{{Input Buffer-Overflow patch.}}
\label{fig:bo-patch}
\end{subfigure}

\begin{subfigure}[t]{\linewidth}
\centering
\includegraphics[width=\linewidth]{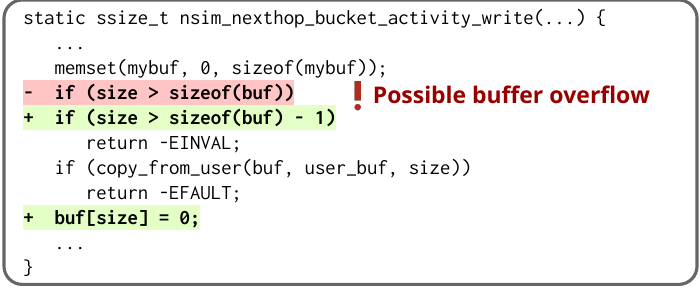}
\caption{\textbf{CVE-2024-50259}, found by the checker for the patch above.}
\label{fig:bo-bug}
\end{subfigure}

\caption{Example vulnerabilities detected by \tech.}
\end{figure}

\parabf{CVE-2025-21715.}
Figure~\ref{fig:uaf-patch} (the input patch to \tech) shows a fix for a Use-After-Free vulnerability.  
In this patch, \CodeIn{free\_netdev} must be invoked only after all the references to its private data, otherwise, it could cause a Use-After-Free issue.
Leveraging this patch, the checker generated by \tech identified a similar bug in \CodeIn{dm9000\_drv\_remove}, as shown in Figure~\ref{fig:uaf-bug}, where \CodeIn{dm} (the private data of \CodeIn{ndev}) remains in use after \CodeIn{ndev} is freed, causing a Use-After-Free.  
This newly discovered issue was assigned CVE-2025-21715.

\parabf{CVE-2024-50259.}
Figure~\ref{fig:bo-patch} shows an input patch fixing a buffer overflow vulnerability.  
The patch mitigates the risk by limiting the number of bytes copied via \CodeIn{copy\_from\_user} to \CodeIn{sizeof(mybuf) - 1}, thereby preserving space for a trailing zero.  
This trailing zero is essential for subsequent string operations, such as \CodeIn{sscanf}, to function correctly.  
Taking this patch as input, the checker generated by \tech identified a similar bug in \CodeIn{nsim\_nexthop\_bucket\_activity\_write}, as shown in Figure~\ref{fig:bo-bug}.  
In this case, the omission of appending a trailing zero after copying data from userspace could lead to improper string handling and potential overflow issues.  
This detected issue was subsequently fixed by adding the trailing zero and was assigned CVE-2024-50259.

\subsection{RQ3: Orthogonality with Smatch}

Since no comparable automated static analyzer generation approaches exist for this domain, we evaluate \tech against expert-written checkers. Our baseline is provided by \smatch~\cite{smatch}, which is widely used in Linux kernel analysis and supports tailored checks for all bug types considered in Table~\ref{tab:collect-commits}.
We conducted the comparative analyses by running \smatch on the entire codebase to determine if it could detect the bugs found by \tech.

\smatch reported a total of 1970 errors and 2870 warnings across the kernel. We manually inspected all the files where bugs were detected to assess whether any of the true positive bugs identified by \tech were also detected by \smatch. Notably, \smatch \emph{failed to detect any of our true positive bugs}, underscoring the unique detection capabilities of \tech. Further analysis of \smatch's checkers revealed that they do not fully leverage the domain-specific knowledge embedded in the Linux kernels---a resource that \tech effectively extracts from historical patches.
For instance, \smatch's \CodeIn{check\_deref} checker employs static range analysis to identify potential null pointers but lacks domain-specific insights. It fails to recognize that functions like \CodeIn{devm\_kzalloc} may return \CodeIn{NULL} under error conditions that conventional static range analysis cannot detect.
Consequently, \smatch identified only three potential null pointer dereferences, all of which were confined to unit test files rarely prioritized by developers.
We conclude that \tech and \smatch detect different classes of bugs, demonstrating \tech's effectiveness in learning domain knowledge from patches and subsequently identifying diverse bugs and vulnerabilities.

\subsection{RQ4: Effectiveness of Components}

\subsubsection{Bug Triage Agent}
\label{subsec:triage}
To evaluate our bug triage agent, from the 39 valid checkers, we sampled up to 5 reports per checker to reduce manual inspection efforts, aiming to reduce manual inspection efforts while maintaining evaluation coverage.
In total, we collected 79 reports from 18 checkers, while the remaining 21 valid checkers didn't generate reports.
Our triage agent classified 29 reports as ``bug'' (positive) and 50 as ``not-a-bug'' (negative). These classifications were compared against ground truth labels established by manual review from two authors. The agent achieved 7 true positives (TP), 22 false positives (FP), 50 true negatives (TN), and \emph{zero false negatives} (FN). The absence of false negatives is particularly important, as it indicates the agent effectively prioritized all potentially true bugs in this set, minimizing the risk of overlooking genuine issues, even though the 22 false positives require further filtering.

\revision{We also evaluated 5-way self-consistency via majority voting~\cite{wang2022self}. For each report, we ran our triage agent five independent times and labeled the report as a “bug” only if the agent made that prediction in at least $t$ of the runs.
Compared to our single-sample baseline, which identified 7 TPs and 22 FPs, majority voting did not offer a significant improvement.
Using a threshold of $t=3$ kept the TP count at 7 but slightly increased FPs to 24. A stricter threshold of $t=4$ also resulted in 7 TPs while reducing FPs to 20.
Ultimately, majority voting only slightly shifted the false positive count in this setting.
There could be two potential reasons. 
For TPs, the agent likely identifies these bugs with high confidence, meaning a single run is sufficient to detect all of them.
For FPs, the agent's predictions are likely less confident and more inconsistent across different runs.}

\subsubsection{Ablation Study}
\label{subsec:ablation}
\begin{table}[t]\centering
\caption{\textbf{Ablation study results.} ``Default'' means \tech's standard configuration utilizing multi-stage synthesis, fixed few-shot examples, and the O3-mini model. Alternative configurations are compared against this baseline.}
\label{tab:ablation}
\small
\begin{tabular}{l|r|rrrr}\toprule
\multirow{2}{*}{Variants} &\multirow{2}{*}{Valid} &\multicolumn{3}{c}{Errors} \\\cmidrule{3-5}
& &Syntax &Runtime &Semantics \\\midrule
\textbf{Default} &12 &28 &0 &75 \\
\midrule
W/o multi-stage &8 &52 &3 &75 \\
W/ RAG &12 &37 &4 & 62 \\
\midrule
W/ GPT-4o &11 &31 &0 &76 \\
W/ DeepSeek-R1 & 11 &29 &8 &66 \\
W/ Gemini-2-flash &4 &130 &2 &44 \\
\bottomrule
\end{tabular}
\end{table}

To evaluate our design choices, we created a sample dataset of patch commits for an ablation study.
We randomly sampled 2 commits from each bug type using zero as the random seed.
This resulted in a dataset of 20 commits (2 commits × 10 bug types).
Table~\ref{tab:ablation} shows the overall results of our ablation study.

\parabf{Checker synthesis.}
First, we assessed the impact of our default \emph{three-stage synthesis approach} (bug pattern analysis, plan synthesis, checker implementation) compared to directly synthesizing checkers in a \emph{single stage} (omitting explicit pattern/plan steps), using identical few-shot examples. As shown in Table~\ref{tab:ablation}, the multi-stage approach proved more effective, yielding valid checkers for 12 commits compared to only 8 for the single-stage method. Furthermore, the single-stage approach resulted in significantly more syntax errors (52 vs. 28), often leading to checkers that failed to compile. This highlights the value of the structured multi-stage process for improving both validity and compilability.

Second, we explored using \emph{Retrieval-Augmented Generation (RAG)}~\cite{gao2023retrieval} for selecting few-shot examples dynamically, comparing it against our default set of \emph{fixed examples}. It utilizes a knowledge base derived from 118 official \csa checkers~\cite{clang-static-analysis}, embedded using \CodeIn{text-embedding-ada-002}~\cite{text-embedding}. During synthesis, three relevant examples are retrieved based on semantic similarity. Our results indicate that RAG-based example selection achieved comparable effectiveness to our fixed examples, also generating valid checkers for 12 commits. However, because the official \csa checkers used are substantially longer than our curated fixed examples, the RAG approach incurred approximately double the input/output token cost. Due to similar effectiveness, our default fixed few-shot examples offer better cost-efficiency for this task.

\parabf{\llm choice.}
We evaluated the performance of \tech across different language models for checker synthesis.
In addition to our default model, O3-mini, we tested GPT-4o, Gemini-2-flash, and open-source DeepSeek-R1.

As detailed in Table~\ref{tab:ablation}, O3-mini yielded the most valid checkers (12). GPT-4o and DeepSeek-R1 performed comparably to each other, generating 11 valid checkers each, only slightly fewer than O3-mini. This suggests that multiple high-capability models, including open-source options, are viable, although minor differences in performance exist.

In contrast, Gemini-2-flash performed substantially worse, producing valid checkers for only 4 commits.
Upon closer inspection, we found that Gemini-2-flash struggled with \csa implementation, frequently using non-existent APIs and generating syntax errors at a much higher rate (130 vs. 28).
This highlights a crucial insight: successful checker synthesis demands more than general coding proficiency; accurate knowledge or inference of the target framework's specific APIs and conventions (\csa in this case) is essential.

\section{\revision{Limitations and Discussion}}

\parabf{\revision{Limitations.}}
\revision{\tech faces challenges with highly complex bug patterns, particularly those involving state-machine reasoning, such as use-after-free, and concurrency issues that require analyzing multi-threaded code and locking schemes.
Synthesizing precise checkers for these issues is difficult for today's \llm{s}, as it requires a sophisticated understanding of a program's temporal and interprocedural behavior.
To address this, a promising future direction is to improve the analysis agent to automatically abstract complex bug reports into more formal state-machine representations. 
This would enable the synthesis of precise, state-aware checkers that can trace the specific sequences of operations leading to a bug.
A second, complementary strategy is to focus on selectively identifying high-quality, canonical bug fixes.
This involves developing methods to filter out patches that are overly complex or specific, allowing \tech to learn from reusable, idiomatic repair patterns and improve the overall performance of checker synthesis.}

\parabf{\revision{Generability.}}
\revision{\tech employs a general three-stage workflow: LLM-driven synthesis of checkers from bug-fix patches, patch-grounded validation, and triage/refinement to reduce false positives.}
\revision{While the Linux kernel served as an ideal and challenging initial target due to its scale, complexity, and importance, our approach is not fundamentally tied to it. The workflow's applicability extends to any project that meets two criteria: a version history with bug-fix commits for learning, and a static analysis framework to serve as a synthesis target (\eg Chromium~\cite{chromium}).}
\revision{Furthermore, the implementation is not limited to C/C++. Although we demonstrated its use with the Clang Static Analyzer~\cite{clang-static-analysis}, the pipeline can be readily adapted—via a small set of few-shot examples—to generate checkers or rules for other ecosystems (\eg CodeQL~\cite{codeql} or Semgrep~\cite{semgrep}).}

\section{Related Work}

\revision{This section mainly discusses related work on both traditional and LLM-based static analysis, contextualizing the contributions of \tech.
While our work focuses on synthesizing \textit{static} analyzers that inspect source code without execution, it is worth noting that checker synthesis has also been explored for generating \textit{dynamic} runtime checks~\cite{t2c, lou2022demystifying,TrainCheckOSDI2025}.
These approaches are fundamentally different from \tech.
Dynamic checkers typically learn rules from test executions~\cite{t2c} to detect bugs based on concrete runtime states, such as capturing semantic ``grey-failures''.
In contrast, \tech learns from source code patches to create static checkers that can cover all potential execution paths, including corner cases rarely triggered during program execution.}

\subsection{Traditional Static Analysis}

Given the cruciality of the Linux kernel and the diversity of its bugs, many static analyzers have been developed. These generally fall into several categories based on their approach.

\parabf{Rule/Model-based analyzers.} A significant body of work focuses on detecting specific classes of bugs using predefined rules or models. Examples include \crix~\cite{kernel-static-lu2019detecting} (detecting missing checks via def-use slices), \goshawk~\cite{kernel-static-lyu2022goshawk} (memory corruption analysis), UBITect~\cite{zhai2020ubitect} (Use Before Initialization bugs), \textsc{CRed}~\cite{yan18uaf} (Use-After-Free detection), LR-Miner~\cite{li2024lr} (data races via locking rules), tools targeting refcounting bugs based on derived anti-patterns~\cite{kernel-static-refcounting} or specific conventions~\cite{liu22linkrid}, DCUAF~\cite{bai19uaf} (concurrent Use-After-Free), and SUTURE~\cite{zhang21suture} (taint analysis for userspace input vulnerabilities). While often effective for their targeted bug classes, these approaches typically require extensive human expertise for analysis design and implementation, limiting their scalability and adaptability to newly emerging bug patterns in the rapidly evolving kernel.

\parabf{\revision{Deviation-based specification inference.}}
\revision{Methods such as~\cite{engler2001bugs,min2015cross,ahmadi2021finding} infer specifications by assuming the \emph{majority} of uses are correct and flagging deviations. While effective in some settings, classic systems rely on a small, fixed set of rule \emph{templates} plus probabilistic clustering/ranking, which constrains bug-pattern coverage and—without strong semantic post-checks or refinement—often yields higher FPR~\cite{engler2001bugs,min2015cross}. FICS~\cite{ahmadi2021finding} reduces template dependence via ML grouping of functionally similar code, but still rests on the majority-correct assumption and similarity heuristics, which can miscluster code and introduce noise.}

\parabf{Patch-based specification inference.}
Another line of work leverages historical patches to \textit{infer specifications}, which are then used by separate checkers. For instance, APHP~\cite{aphp} extracts API Post-Handling (APH) specifications from both code and descriptions in patches to detect APH violations. A very recent work, Seal~\cite{seal}, analyzes security patches to infer diverse specifications for Linux interfaces.
While powerful, these methods generally rely on existing static analysis infrastructure to enforce the inferred specifications.
We attempted to include Seal in our experimental comparison, but encountered practical difficulties as its publicly available version depends on private commercial tools and was not compatible with the recent Linux kernel versions.

\parabf{General static analysis frameworks.} Other systems focus on improving underlying static analysis techniques for better precision or efficiency across multiple bug types. FiTx~\cite{suzuki2024fitx}, for example, implements fast analysis for single compilation units, while PATA~\cite{li22pata} enhances alias analysis precision using path information. These systems often encode bug patterns as state machines~\cite{suzuki2024fitx, li22pata}, but designing these patterns still typically involves manual effort.

\parabf{Our synthesis approach.} In contrast to the above, \tech uses \llm{s} to \textit{synthesize the static analyzer itself} directly from historical patch information. Instead of relying on predefined rules or inferring specifications for existing checkers, \tech learns both the bug pattern and the corresponding detection logic, generating a new checker automatically. This approach fundamentally differs by automating the analyzer creation process, potentially offering greater adaptability to new bug types compared to traditional methods or specification inference. This distinction may also make \tech more versatile; while specification-inference tools excel at finding bugs related to the interfaces they model, \tech can synthesize checkers for a broader range of issues demonstrated in patches, including problems not strictly tied to interface misuse, such as certain integer overflows or complex logic errors. Furthermore, although \tech currently generates \csa checkers, the core idea of LLM-based checker synthesis is potentially generalizable to target other static analysis frameworks, given the broad training data of modern \llm{s}. Thus, \tech complements existing approaches by providing a mechanism to automatically generate analyzers for newly observed bug patterns found in patches.

\subsection{LLM-Based Static Analysis}

Recent LLM advancements enable new static analysis techniques, differing significantly from our synthesis approach.

\parabf{LLM-augmented static analysis.} Some techniques use LLMs to \textit{assist} existing tools by automating tasks previously requiring manual effort.
\revision{For example, LLMs can infer taint specifications for external APIs (IRIS~\cite{li2024llm}, Artemis~\cite{ji2025artemis})},
generate post-constraints to prune analysis paths (\llift~\cite{li2024enhancing}), or infer resource-handling intentions (\inferroi~\cite{wang2024resourceleak}). While helpful, these methods still fundamentally depend on a substantial, human-developed analyzer core\revision{, \eg the manually defined rules for taint propagation~\cite{ji2025artemis}}.
This can limit their ability to easily generalize to detect novel bug types beyond those originally targeted by the core analyzer's design.

\parabf{Direct code analysis with LLMs.} Other work uses LLMs to \textit{directly analyze} source code, using techniques like Retrieval-Augmented Generation (\vulrag~\cite{du2024vul}), prompt engineering~\cite{cy24prompt}, or fine-tuning~\cite{shestov}. However, directly applying LLMs to scan large systems like the Linux kernel is often prohibitively expensive and faces scalability challenges.

\parabf{Our synthesis approach.}
\revision{In contrast, \tech uses LLMs to \textit{synthesize the entire static analyzer} from patches. Unlike LLM-augmented tools, this minimizes reliance on pre-existing manual analyzer development, enhancing adaptability for diverse bug types.
Unlike direct LLM analysis, \tech generates efficient, reusable static checkers, avoiding the high costs and scalability issues of scanning massive codebases directly with LLMs.
A very recent concurrent work, MoCQ~\cite{mocq}, also explores checker/query synthesis, but it focuses on general bug patterns and relies on manual examples for validation.
\tech distinguishes itself not only by automatically inferring specific, nuanced patterns from patches but also by incorporating a closed-loop triage-refinement pipeline to iteratively improve checker precision.
This fully automated refinement process makes our approach more scalable for detecting complex defects in system software, establishing it as a practical paradigm for applying LLM intelligence to large-scale static analysis.}

\section{Conclusion}

This paper introduces \tech, a novel approach that transforms how LLMs can contribute to static analysis for complex systems like the Linux kernel. By synthesizing static analyzers rather than directly analyzing code, \tech bridges the gap between LLMs' reasoning capabilities and the practical constraints of analyzing massive systems.
\tech's practical impact is shown by the discovery of \numBug new, long-latent bugs in the Linux kernel, with \numConfirmed confirmed, \numFixed fixed, and \numCVE CVE assigned.

Looking forward, \tech opens new possibilities for scalable \llm-based static analysis. Future work could extend this approach to other systems beyond the Linux kernel, incorporate additional learning paradigms, and further refine checker generation techniques to address more complex bug patterns. By leveraging \llm{s} to synthesize tools rather than perform analysis directly, we establish a scalable, reliable, and traceable paradigm for utilizing AI in critical software security applications.

\section*{Acknowledgments}

We are grateful to the anonymous reviewers for their valuable feedback that helped to improve this paper. This work was partially supported by NSF grant CCF-2131943.

{\footnotesize \bibliographystyle{acm}
\balance
\bibliography{main}}

\end{document}